\documentclass[reprint,superscriptaddress,prl,amsmath,amssymb]{revtex4-2}

\usepackage{graphicx}
\usepackage{dcolumn}
\usepackage{bm}
\usepackage{xcolor}
\usepackage{siunitx}
\usepackage{relsize}
\usepackage{subfiles}
\usepackage{microtype} 

\newcommand{\tuwien}{Institute for Theoretical Physics, TU Wien, Wiedner Hauptstra\ss e 8-10/136, A-1040 Vienna, Austria}
\newcommand{\ATI}{Vienna Center for Quantum Science and Technology, Atominstitut, TU Wien, Stadionallee 2, A-1020 Vienna, Austria}

\newcommand{\comment}[1]{}

\begin{document}

\preprint{APS/123-QED}
\title{Triggered Superradiance and Spin Inversion Storage in a Hybrid Quantum System}

\author{Wenzel Kersten}
\email{wenzel.kersten@tuwien.ac.at}
\affiliation{\ATI}
\author{Nikolaus de Zordo}
\affiliation{\ATI}
\author{Oliver Diekmann}
\author{Tobias Reiter}
\author{Matthias Zens}
\affiliation{\tuwien}
\author{Andrew N. Kanagin}
\affiliation{\ATI}
\author{Stefan Rotter}
\affiliation{\tuwien}
\author{J{\"o}rg Schmiedmayer}
\affiliation{\ATI}
\author{Andreas Angerer}
\affiliation{\ATI}

\date{\today}

\begin{abstract}
We study the superradiant emission of an inverted spin ensemble strongly coupled to a superconducting cavity. After fast inversion, we detune the spins from the cavity and store the inversion for tens of milliseconds, during which the remaining transverse spin components disappear. Switching back on resonance enables us to study the onset of superradiance. A weak trigger pulse of a few hundred photons shifts the superradiant burst to earlier times and imprints its phase onto the emitted radiation. For long hold times, the inversion decreases below the threshold for spontaneous superradiance. There, the energy stored in the ensemble can be used to amplify microwave pulses passing through the cavity.
\end{abstract}

\maketitle

Superradiance is the process by which an ensemble of excited two-level systems synchronizes to produce a short, highly coherent burst of light \cite{dicke1954coherence}. The build-up of correlations during the collective decay, mediated by an enhanced coupling to a common mode, gives rise to non-linear scaling of the decay rate with the number of emitters \cite{superradiance}. Superradiant (SR) emission is not only fundamental to many fields of physics, but also attracts increasing interest for applications in metrology \cite{koppenhofer2022dissipative}, laser physics \cite{bohnet2012steady,zhang2021ultranarrow,wu2022superradiant} and quantum technology in general \cite{kuzmich2003generation,yang2021realization,kim2018coherent,pennetta2022observation,araujo2016superradiance,kim2022photonic,sherman2022diamond}. SR phenomena are at the heart of the transition from a genuine quantum regime, where individual fluctuations of the vacuum field will jump-start the collective decay of inverted emitters, to the classical regime, where the emission is akin to that of a macroscopic radiating dipole. 

Whereas experiments on superradiance have recently been successfully transferred from atomic ensembles to solid-state spin systems \cite{angerer2018superradiant,quach2022superabsorption}, the possibilities this opens up for controlling and exploiting superradiance for applications have been very little explored so far. Progress in this direction has primarily been hindered by the fact that systems giving rise to superradiance are fundamentally unstable, reacting to the slightest disturbance. While this extreme sensitivity even to weak signals poses a great challenge for experimental implementation, it also provides potential avenues for applications in sensor and detector technology \cite{koppenhofer2022dissipative,RAMII}.

Our work is enabled by an experimental platform that allows us {\em(i)} to invert a large ensemble of nitrogen-vacancy (NV) spins, and {\em(ii)} to hold and stabilize the stored inversion for up to \SI{20}{\milli \second} – four orders of magnitude longer than the timescale of the SR emission. Stabilization is achieved by rapidly detuning the spins from cavity resonance after their inversion, switching off the interaction with the mode. This allows us to study and control the emission of a SR burst that releases the energy stored in the ensemble. We employ weak microwave (MW) pulses to trigger the SR emission and also explore a regime with reduced inversion, where the spins act as a gain medium.

\begin{figure*}
    \centering
    \includegraphics[width=180mm]{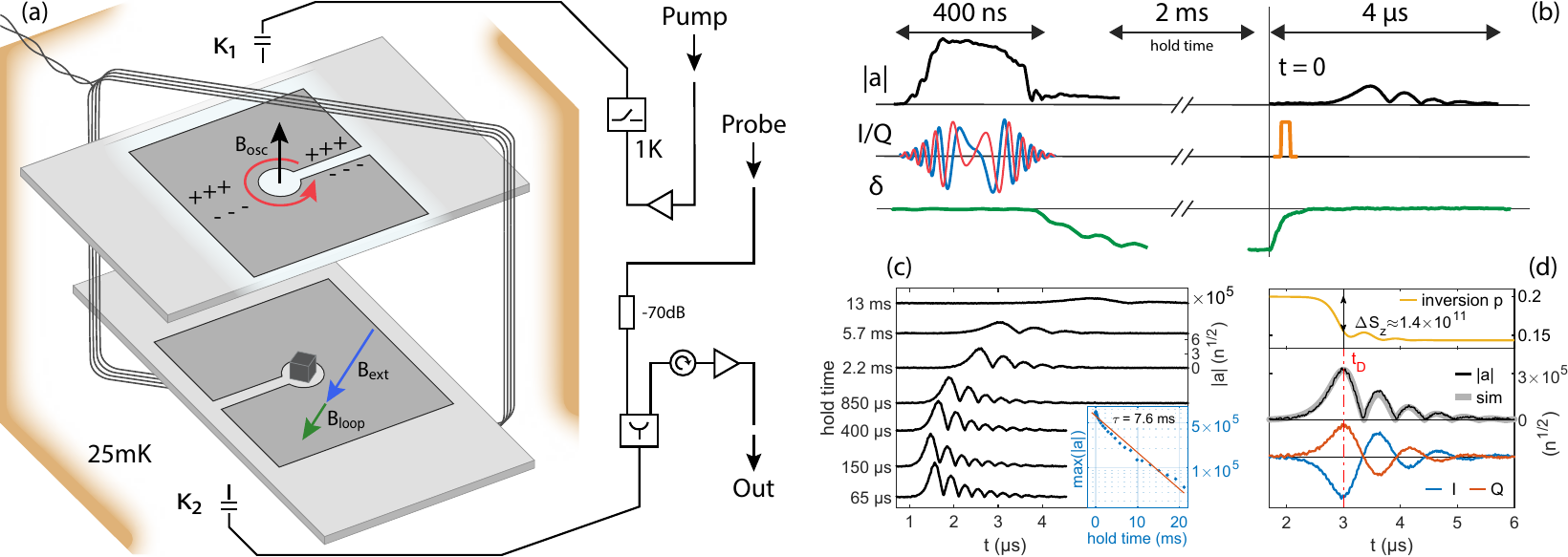}
    \caption{(a) Schematic of the MW cavity located in a dilution refrigerator at $\SI{25}{\milli \kelvin}$ and connected to a homodyne MW setup. Two sapphire chips with opposing split ring structures and the diamond sample are stacked inside a copper box. Between the center holes the oscillating magnetic field homogeneously penetrates the sample. A superconducting wire loop wrapped around the chips enables rapid spin detuning. Port 1 is connected to the pump-line, which can be decoupled at the $\SI{1}{\kelvin}$ stage using a solenoid switch for noise suppression. Port 2 is connected to the out-line for acquiring data, and the attenuated probe-line for injecting weak trigger pulses. (b) Experiment sequence: we use a modified chirp pulse (red/blue) to invert the spin ensemble and subsequently modulate the spin detuning $\delta$ to store the inversion. After a variable hold time we bring the spins back into resonance and measure the cavity amplitude $\lvert a \rvert$ of the SR decay, optionally triggered by a short probe pulse (orange). (c) SR decays for varying hold times, triggered by the pump-line amplifier noise. The inset shows the SR decay maxima with an exponential fit in a semi-log plot. (d) Example data and simulation of an SR decay and its quadratures $I / Q$, with simulated inversion $p$. The vertical line indicates $t_\mathrm{D}$, the time of maximum cavity amplitude. The number of cavity photons $n = \lvert a \rvert^2$ (calibration in Supplemental Material) agrees well with the estimated number of decaying spins $\Delta S_z$.}
    \label{fig1:setup}
\end{figure*}

Our resonator [see Fig.~\ref{fig1:setup}(a)] is based on two opposing superconducting chips that exhibit a small mode volume with homogeneous coupling strength, while retaining a high quality factor of $Q \approx 3000$. This design allows us {\em(i)} to reach the regime of strong collective spin-cavity coupling already with a number of NVs that is reduced by three orders of magnitude and {\em(ii)} to add a small loop of superconducting wire which enables magnetic tuning of the spins in and out of the cavity resonance faster than the SR timescale.  Previous cavity realizations for this type of SR system ~\cite{angerer2016collective, ball2018loop} would resist a sudden field change due to induced currents in their bulk structures.

The two sapphire chips with a $\SI{200}{\nano \meter}$ thin layer of $16 \times \SI{16}{\milli \meter \squared}$ Niobium are mounted in a copper housing. The identical patterns on both chips feature a hole in the center from which a \SI{4}{\micro\meter} slit reaches outwards, resembling a split ring resonator \cite{splitringres}. The chips are stacked, with the roughly cube-shaped diamond sample placed between the center holes. The hole radii, the distance between the chips and the sample size are all of similar dimension $d \sim \SI{200}{\micro \meter}$. This configuration results in a resonance frequency of $\omega_\mathrm{c}/2 \pi = \SI{3.105}{\giga \hertz}$ and linewidth of $\kappa/ 2\pi = \SI{0.51}{\mega \hertz}$ (HWHM). 

The resonator couples homogeneously to all spins with a single spin coupling strength of $g_0 \approx \SI{2}{\hertz}$, resulting in a collective coupling of $g_{\mathrm{coll}}/2\pi = \SI{5.17}{\mega\hertz}$ for a total number of spins of $N \approx \num{6.4e12}$. The spin system's coherent response is determined by the effective ensemble linewidth of $\Gamma_\perp/ 2\pi = \SI{4.27}{\mega \hertz}$ \cite{molmer1}, combining the inhomogeneously broadened spin frequency distribution \cite{sandner2012strong} and the individual spin's linewidth of $\gamma_\perp/2\pi \approx \SI{208}{\kilo \hertz}$ \cite{putz2017spectral}. More details on the resonator and theoretical treatment can be found in the Supplemental Material. The resulting cooperativity parameter of our coupled system is $C = g_{\mathrm{coll}}^2 /(\kappa \Gamma_\perp) \approx 12.2$.

To begin our explorations, we magnetically tune all 4 NV sub-ensembles into resonance with the cavity using a static vector field and a loop current of $\SI{1}{\ampere}$. The NVs are initially prepared in a state close to the ground state and act as effective two-level sytems. Next, we use a \SI{400}{\nano \second} modified chirp pulse with a Gaussian envelope to invert the spins. We then rapidly switch off the loop current in about $\SI{200}{\nano \second}$ using a semiconductor switch, detuning the spins by $\delta/2\pi \approx \SI{26}{\mega \hertz}$ for a given hold time. This detuning by more than the ensemble linewidth inhibits the SR interaction of the spin ensemble with the cavity mode \cite{molmer1}, thereby storing the inversion. Initially, the stored population in the upper spin state is $\approx 67\%$. For details of the initialization see Supplemental Material.

During the hold time, the remaining transversal component of the collective spin vector $S_- = S_x - \mathrm{i}S_y$, which initially persists after the creation of the partially inverted state, undergoes dephasing and is effectively eliminated. When tuning the ensemble back into resonance, we thus create a metastable inverted state whose tipping angle $\theta = \arctan(\lvert S_-  \lvert /  S_z )$ with respect to the $z$-axis in the Bloch sphere is exponentially decreased for longer hold times. If the product of the stored ensemble inversion $-1 \leq p \leq +1$ and cooperativity is above the threshold $p C > 1$, this metastable state will become unstable and decay by emitting a SR photon burst, as shown in \cite{molmer1}. Here, the inversion parameter $p$ is implicitly defined by $ S_z = \frac{1}{2} \langle \sum_j  \sigma_z^j \rangle = p N/2$. In this state, the presence of even a single photon in the cavity will stimulate the collective emission of radiation, starting a self-accelerating photonic avalanche. During this process, the energy released in the form of cavity photons gradually builds up, reaches a maximum and then oscillates back and forth between the two subsystems, before the process stops due to the dephasing of the spins and their decoherence. The full experimental sequence is summarized in Fig.~\ref{fig1:setup}(b).

\begin{figure*}[t]
    \includegraphics[width=180mm]{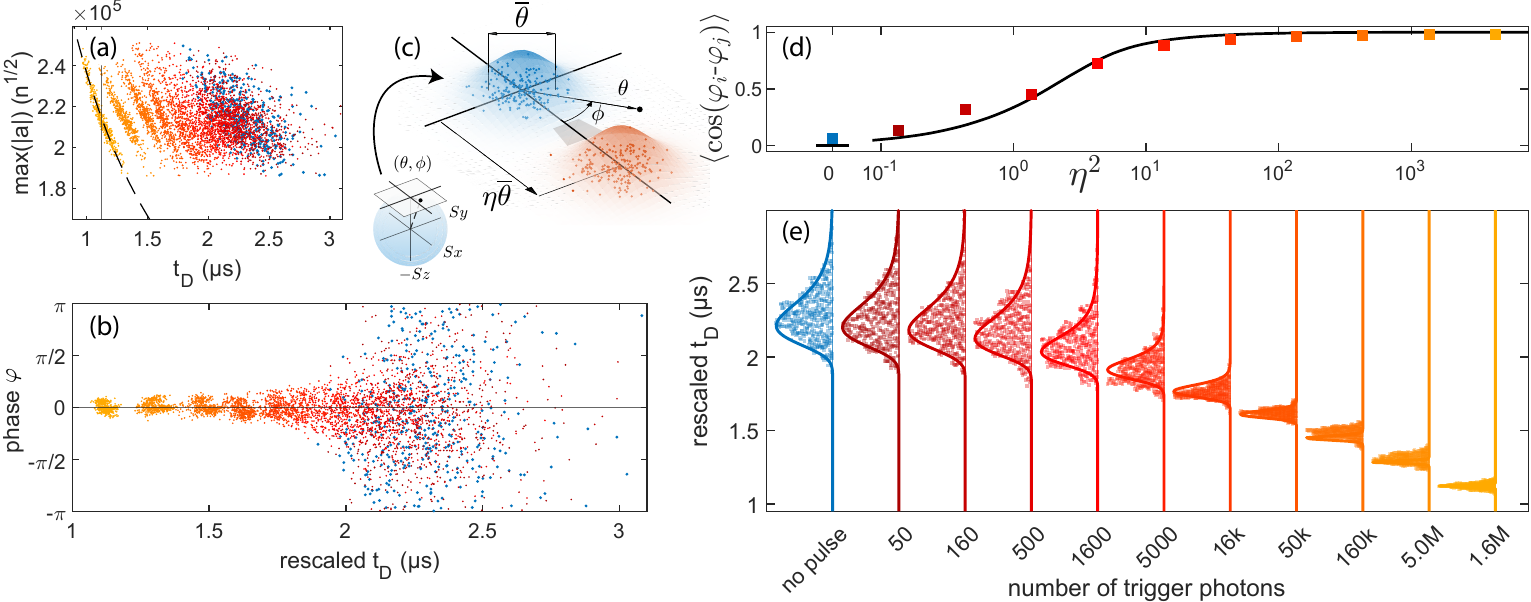}
    \caption{Triggering the SR decay with $\SI{100}{\nano \second}$ pulses containing different numbers of photons, color-coded according to (e). (a) Maxima of the SR decay amplitudes plotted over delay times $t_\mathrm{D}$. (b) Corresponding SR decay phases $\varphi$ plotted over a rescaled $t_\mathrm{D}$ axis. The rescaled $t_\mathrm{D}$ values result from a transformation that aligns the dashed curve in (a) onto the vertical line. (c) Initial state of the collective spin vector with coordinates $(\theta,\phi)$ close to the north pole of the Bloch sphere: in-plane distribution before (blue) and after (red) the coherent displacement $\eta$ in units of the width $\overline{\theta}$ induced by the trigger pulse. (d) Phase average over all runs $\langle \cos (\varphi_i$-$\varphi_j) \rangle$ quantifying the phase randomness from the measured sets of $\varphi$. (e) Swarm plots of the delay time $t_\mathrm{D}$ data. The solid lines in (d) and (e) are obtained from our theoretical description, varying only the parameter $\eta$.}
    \label{fig2:decay_statistics}
\end{figure*}

Our first notable result is presented in Fig.~\ref{fig1:setup}(c), where we plot the SR decay pulses for varying inversion hold times. Here, the SR decay is triggered by noise from the high power amplifier of the pump-line. The measured SR dynamics are captured in a semi-classical description using the Maxwell-Bloch equations \cite{carmichael1999statistical}. We model the time evolution starting from an inverted state with a slight tipping angle accounting for fluctuations that initiate the SR decay (see Supplemental Material). To simulate the measured signals of $\lvert a \rvert$ we only adjust the ensemble inversion $p$ and a time offset, resulting in curves as shown in Fig.~\ref{fig1:setup}(d). The role of fluctuations at the start of the SR decay process is studied in more detail below. We find the decay maximum $\max(\lvert a \rvert)$, an indirect measure of the energy stored by the spins, to decrease roughly exponentially with increasing hold times, exhibiting a characteristic timescale of $\tau = \SI{7.6}{\milli \second}$ [see inset Fig.~\ref{fig1:setup}(c)]. For hold times longer than $\SI{20}{\milli \second}$, the inversion has already decreased below the threshold $pC = 1$ for spontanoeus superradiance. We propose two timescales for the relaxation of the inverted state. First, on a millisecond timescale, the ensemble is rapidly randomized due to spin-spin interactions involving NVs with short lifetimes (so called {\it fluctuators} \cite{choi2017depolarization}), acting as local sinks for the inversion via spin diffusion. Second, when $p=0$ is reached, the ensemble relaxes to the ground state on a longer timescale, characterized by $T_1 = \SI{134}{\second}$ (see Supplemental Material).

We now focus on the onset of the SR decay process and the possibility to trigger it prior to its self-decay. Using a $\SI{2}{\milli \second}$ hold time, we give the cavity mode enough time to reach thermal equilibrium after the inversion pulse and subsequent decoupling from the high power amplifier noise by the solenoid switch, with an estimated number of $\overline{n}\approx 3$ thermal photons remaining. The partially inverted state that is brought back into resonance has zero tipping angle apart from unavoidable quantum and thermal fluctuations. Another $\SI{150}{\nano \second}$ after switching back the detuning current (defined as $t = 0$), we send a $\SI{100}{\nano \second}$ trigger pulse through the highly attenuated MW probe-line. The pulse is resonant with the cavity and contains a calibrated number of photons (see Supplemental Material). The experiment is repeated many times for varying numbers of trigger photons, and without trigger pulse. For every run, we extract the delay time $t_\mathrm{D}$ and the $I_\mathrm{D}/Q_\mathrm{D}$ quadrature values of the SR decay maximum [cf. Fig.~\ref{fig1:setup}(d)]. 
The SR decay amplitudes $\max(\lvert a \rvert) = \sqrt{I^2_\mathrm{D} + Q^2_\mathrm{D}}$ show variations of $\pm 10 \%$ between runs as visible in Fig.~\ref{fig2:decay_statistics}(a), mainly caused by the solenoid switch's latching mechanism. To clearly study the influence of the number of trigger photons on the delay times, we adjust for the expected systematic dependence of $t_\mathrm{D} \propto \max(\lvert a \rvert)^{-1}$ and rescale the $t_\mathrm{D}$ data. The SR decay phases $\varphi = \arctan(Q_\mathrm{D}/I_\mathrm{D})$ are independently corrected for a linear phase drift with $t_\mathrm{D}$, caused by a minor constant detuning of the spins. Details of both methods are given in the Supplemental Material. The resulting sets of phases and rescaled delay times are presented in Figs.~\ref{fig2:decay_statistics}(b) and \ref{fig2:decay_statistics}(e).

Clearly, stronger trigger pulses with higher numbers of photons $n_\mathrm{trig}$ lead to earlier $t_\mathrm{D}$ values and narrower distributions for $t_\mathrm{D}$ and $\varphi$. While our simulation allows to describe the decay process starting from a slightly tipped initial collective spin vector, it is the randomness in the initial conditions that leads to the observed variance in time and phase. These thermal and quantum fluctuations are not included in our semi-classical model. To understand the observed phenomena, we split the analysis of the SR decay into two stages \cite{superradiance, RAMI, RAMII}. 

The decay process starts with a linear stage, in which the (optional) trigger pulse leads to a coherent rotation of the collective spin vector about an axis defined by the phase of the pulse, which is kept identical for all runs. Prior to this rotation, the initial state is located very close to the $+z$-axis but with a small tipping angle $\theta = \arctan(\lvert S_- \lvert /  S_z )$ and random polar angle $\phi = \arg( S_- )$. As $\cos \theta \simeq 1$ throughout the linear phase, we can treat the spin vector to be confined to a plane with a $z$-offset corresponding to the initial inversion. The geometric construction of this plane is illustrated in Fig.~\ref{fig2:decay_statistics}(c), mathematical formulas of the distribution functions are given in the Supplemental Material. The initial state of the spin vector follows a two dimensional Gaussian distribution of width $\overline{\theta}$ centered at $\theta=0$. The influence of the trigger pulse then causes a displacement in the plane, which we choose to be in the direction of $\phi = 0$. The parameter $\eta$ expresses the displacement in units of the width parameter $\overline{\theta}$. For growing $\eta$, i.e., higher trigger pulse powers, the initially randomly distributed polar angles become increasingly well defined and approach a narrow distribution around $\phi = 0$ [see Fig.~\ref{fig2:decay_statistics}(b)].

After this linear stage, where the collective spin vector is coherently displaced from its random in-plane starting position, we enter a nonlinear regime. Now the SR dynamics dominate and via a collective process of stimulated emission the spin vector accelerates its rotation towards the equator while emitting a considerable burst of MW radiation.

The phase $\varphi$ of the emitted decay pulse is directly determined by the value of $\phi$ at the start of the nonlinear stage. Less directly, we can infer the initial tipping angles $\theta$ from the delay times $t_\mathrm{D}$, which result via the relation \mbox{$t_\mathrm{D} = - 2 T_\mathrm{R} \log\left(\theta/2\right)$} \cite{RAMII}. Here, the parameter $T_\mathrm{R}$ represents the timescale for the SR emission process (see Supplemental Material). With this relation, and the displaced Gaussian distribution that describes $\theta$ and $\phi$ depending on $\eta$ [see Fig.~\ref{fig2:decay_statistics}(c)], we can reproduce the $t_\mathrm{D}$ data in Fig.~\ref{fig2:decay_statistics}(e) and the phase randomness quantified by $\langle \cos (\varphi_i$-$\varphi_j) \rangle$ \cite{RAMII} in Fig.~\ref{fig2:decay_statistics}(d). To this end, we fix the values of the SR timescale $T_\mathrm{R} = \SI{142}{\nano \second}$ and width of the Gaussian $\overline{\theta} = \num[]{5.85e-04}$, and vary only $\eta$.

As the displacement $\eta$ is caused by the MW magnetic field of the trigger pulse, its square is a measure of the energy imparted onto the spin system during the linear stage of the SR process. We can therefore use the $x$-axes in both Figs.~\ref{fig2:decay_statistics}(d,e) interchangeably, confirming $n_{\mathrm{trig}} \propto \eta^2$. Remarkably, a weak MW pulse on the order of $10^{-11}$ photons per spin is sufficient to have an observable effect on the SR decay. By reducing the number of spins while maintaining a high cooperativity, the sensitivity to both amplitude and phase could be further enhanced.

\begin{figure}
    \centering
    \includegraphics[width=85mm]{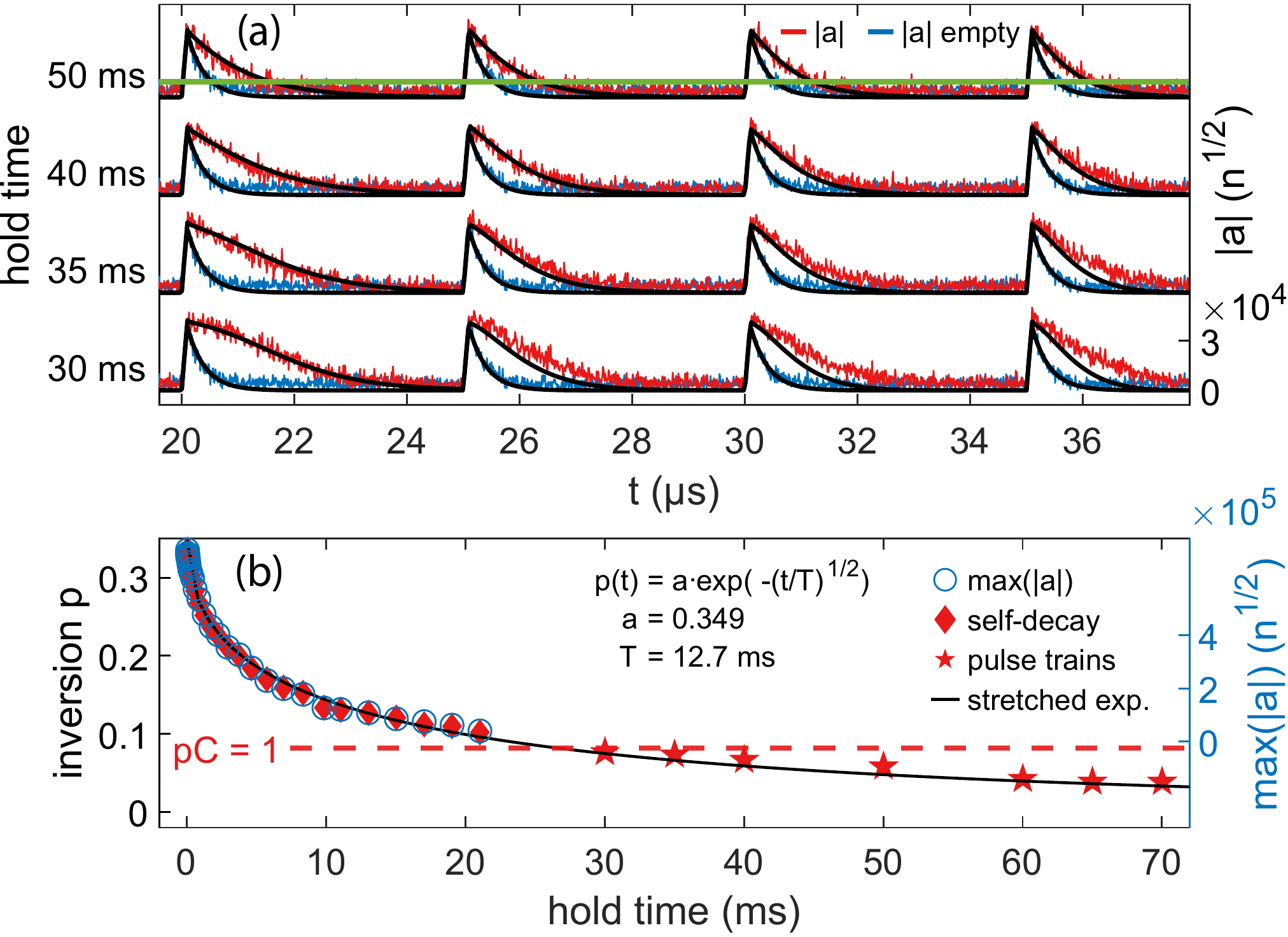}
    \caption{(a) Cavity amplitude $\lvert a \rvert$ for a series of $\SI{100}{\nano \second}$ pulses, each injecting $n_{\mathrm{trig}} \approx \SI{1.5e9}{}$ photons, amplified by the partially inverted spin ensemble in the reduced effective cooperativity regime $pC < 1$ for different hold times (red). In comparison, we plot the signal obtained with an empty cavity where spins are far detuned (blue). For choosing the parameters in our semi-classical model (black), we ignore noise below a certain threshold (green line at the top). (b) Ensemble inversion as a function of hold time, extracted by simulations in the two regimes above and below $pC = 1$. Above this threshold, the pulse maxima (right $y$-axis) follow the values of $p$ from simulations of the self-decays shown in Fig.~\ref{fig1:setup}(c). A stretched exponential with exponent $1/2$ (characteristic for spin diffusion in three dimensions \cite{choi2017depolarization}) is fitted to the inversion.}
    \label{fig3:pulse_trains_inversion}
\end{figure}

We now investigate a regime of reduced effective cooperativity $pC < 1$, where SR emission does not occur spontaneously \cite{molmer1}. To that end, we employ hold times longer than $\SI{20}{\milli \second}$, thus reducing the polarization below the threshold for the SR decay. We probe the system by injecting, at $\SI{5}{\micro\second}$ intervals, a sequence of resonant MW pulses of $\SI{100}{\nano \second}$ duration via the pump-line. Interestingly, in Fig.~\ref{fig3:pulse_trains_inversion}(a), we find that this results in an amplification of the pulses as compared to the empty cavity response (with far detuned spins). Although no spontaneous SR decay occurs on its own, it is still possible to repeatedly extract energy from the stored inversion. The incident pulses hereby effectively supply the necessary coherence that is otherwise constituent to the SR emission, but hindered from building up when the stored inversion is insufficient. Notably, tens of injected MW pulses can be amplified in succession (see Supplemental Material). We are able to replicate the measured dynamics using our numerical model with only the amplitude of the incident pulses (kept fixed for all fits) and the ensemble inversion $p$ as free parameters. These results are combined in Fig.~\ref{fig3:pulse_trains_inversion}(b) with the $p$ values attained by simulating the SR self-decays [cf. Fig.~\ref{fig1:setup}(d)]. The semi-classical model seamlessly captures the behavior of our system in both regimes of high and low effective cooperativity.

In summary, we present an experimental platform to store the energy of an inverted spin ensemble for tens of milliseconds and to release it in a strong SR burst. By initializing the system to a fully upright inverted state, we demonstrate a high sensitivity to weak MW pulses that strongly influence the subsequent SR dynamics via both amplitude and phase of the trigger pulse. The decrease of inversion over time lets us explore a regime of reduced cooperativity without spontaneous SR emission, where the inverted spins effectively act as a gain medium for a series of short MW pulses. Our observations provide insight into the collective behavior of inverted spin systems and its experimental control.
\\

\begin{acknowledgments}
We thank Johannes Majer for discussions and technical support in the initial phases of the project. We acknowledge support by the Austrian Science Fund (FWF) projects I3765 (MICROSENS), P34314 (Spins in Quantum Solids) and P32300, and by the European Union’s Horizon 2020 research and innovation programme (FET-OPEN project FATMOLS, Grant No. 862893) as well as by the Studienstiftung des Deutschen Volkes.
\end{acknowledgments}

\newpage

\onecolumngrid
{\centering
{\large \bf Supplemental Material:} \\ 
\vspace{2mm}
{\normalsize \bf Triggered Superradiance and Spin Inversion Storage in a Hybrid Quantum System} \\
\vspace{7mm}
}
\twocolumngrid

\pagenumbering{arabic}
\renewcommand*{\thepage}{S\arabic{page}}

\renewcommand{\thefigure}{S\arabic{figure}}
\setcounter{figure}{0}

\renewcommand{\theequation}{S\arabic{equation}}
\setcounter{equation}{0}

\section{System Hamiltonian, equations of motion and numerical modelling}
Our system is described by the driven Tavis-Cummings Hamiltonian in the rotating frame \cite{sandner2012strong},
\begin{equation}
    \begin{split}
    \mathcal{H} =& \hbar \, \Delta_\mathrm{c} a^\dagger a + \frac{\hbar}{2} \sum_{j} \Delta_\mathrm{s}^j \sigma_z^j \\
   & +\hbar \, \sum_{j} g_0 \left(a^\dagger \sigma_-^j + \sigma_+^j a\right)  + \mathrm{i} \hbar \, \eta \left( a^\dagger - a \right) ,
    \end{split}
\end{equation}
with $a^\dagger$ ($a$) being the creation (annihilation) operator of the cavity mode and $\sigma_z^j$, $\sigma_\pm^j$ being the Pauli-$z$ and raising/lowering operators for the $j^\mathrm{th}$ spin, respectively. The spins are coupled to the cavity with constant coupling $g_0=g_\mathrm{coll}/\sqrt{N}$, where $g_\mathrm{coll}$ is the collective coupling strength and $N$ is the number of spins. The spin detunings $\Delta_\mathrm{s}^j = [\omega_\mathrm{s}^j + \delta(t)] - \omega_\mathrm{p}$ account for the inhomogeneous broadening of the spin ensemble and the additional shift $\delta(t)$ caused by the detuning loop, while $\Delta_\mathrm{c} = \omega_\mathrm{c} - \omega_\mathrm{p}$ is the detuning of the cavity mode. Both detunings are calculated with respect to the driving frequency $\omega_\mathrm{p}$. Further, the amplitude of the driving field is determined by $\eta$.\\
Using a Lindblad master equation, we take into account the loss rate $\kappa$ for the cavity mode and the spin decoherence rate $\gamma_\perp$, thus yielding a set of coupled equations describing the dynamics of the operators,
\begin{align}
     \dot{a} &= -(\mathrm{i}\Delta_\mathrm{c}+\kappa) a - \mathrm{i} \sum_{j} g_0 \sigma_-^j + \eta\,, \label{eq_MBE_a}\\
     \dot{\sigma}_-^j &= -(\mathrm{i}\Delta_\mathrm{s}^j+ \gamma_\perp) \sigma_-^j + \mathrm{i}g_0 a  \sigma_z^j \,,\label{eq_MBE_sigma_m} \\[5pt]
     \dot{\sigma}_z^j &=  2\mathrm{i} g_0 \left(  a^\dagger \sigma_-^j - a \sigma_+^j   \right)\,.\label{eq_MBE_sigma_z}
\end{align}
Identifying these operators with their expectation values, effectively neglecting correlations between individual spins and the cavity by separating higher order moments into products of their first order counterparts, yields the well known Maxwell-Bloch equations. These equations of motion represent a semi-classical description of our system's dynamics, which can be solved numerically, e.g., to model the response to external stimuli. In our analysis, this description only fails to capture the stochastic nature of the inverted state's initial conditions.

To perform the numerical simulations, we approximate the spin frequency distribution $\rho(\omega)$, which is quasi-continuous due to the large number of spins, by sampling it at 1500 frequencies $\omega_j$ with equidistant spacing $\Delta\omega$. The resulting weights $\rho_j=\rho(\omega_j)\Delta\omega$ are then used to calculate the number of spins, $n_j=\rho_jN$, for each frequency bin $\omega_j$. Assuming identical initial conditions for all spins of one bin, also their dynamics according to Eqs.~\eqref{eq_MBE_sigma_m} and \eqref{eq_MBE_sigma_z} are identical, thus tremendously reducing the number of relevant equations.

We determine the parameters describing our system, namely $\omega_\mathrm{c}$, $\kappa$, $g_\mathrm{coll}$, and $\Gamma_\perp$ [combining $\gamma_\perp$ and the spin frequency distribution in one parameter, see Eq. \eqref{eq:Gamma_perp}] by fitting the steady state solution of the Maxwell-Bloch equations to the transmission signals [c.f. Fig.~\ref{fig:vnaT1}(a)], obtained with a vector network analyzer (VNA). 

For modelling SR emission dynamics on short timescales, we can safely neglect $T_1$ processes, i.e., we do not include these in the master equation.

\section{Spin frequency distribution}
To model the inhomogeneously broadened spin distribution, we fix $\gamma_\perp/2\pi = 1/T_2 = \qty[]{208}{\kilo \hertz}$, where $T_2 = 4.8 \pm \qty{1.6}{\micro \second}$, and the shape parameter $q = 1.39$ of the $q$-Gaussian \cite{sandner2012strong} function $\rho(\omega)$, with both values as reported in \cite{putz2017spectral} for a similar NV diamond sample.
The effective linewidth $\Gamma_\perp = 1/T_2^*$, present in the definition of the cooperativity $C = g_\mathrm{coll}^2/\kappa\Gamma_\perp$ in the main text, can be calculated using \cite{molmer1}
\begin{equation}
    \Gamma_\perp = \left[\int_{-\infty}^{+\infty} \frac{\rho(\omega) d\omega}{\gamma_\perp + \mathrm{i}(\omega - \omega_0)} \right]^{-1},
    \label{eq:Gamma_perp}
\end{equation}
with $\omega_0$ being the spin center frequency, together with the value $\gamma_q/2\pi = \qty{11.0}{\mega \hertz}$ for the FWHM of the q-Gaussian, which is chosen to fit the steady state ground state transmission on resonance, see Fig.~\ref{fig:vnaT1}(a).

\section{Initialization procedure and inversion pulse}
The external magnetic field orientation is chosen to tune all four sub-ensembles into resonance with the cavity with an external field strength of $B_\mathrm{ext} \approx \qty{8}{\milli \tesla}$ created using a three dimensional Helmholtz coil setup, while the field created by the current loop is $B_\mathrm{loop} \approx \qty{1.1}{\milli \tesla}$.
As the third level of the NV ground state manifold is far away from resonance, we can treat the NVs as effective two-level systems. For all experiments, the spins are initialized close to their ground state by waiting $\qty[]{3}{\minute}$ after repeatedly sweeping a MW tone across the cavity resonance for $\qty[]{30}{\second}$, thereby creating a repeatable initial state, reducing wait times between different runs. \\

\begin{figure}[h]
    \centering
    \includegraphics[width=65mm]{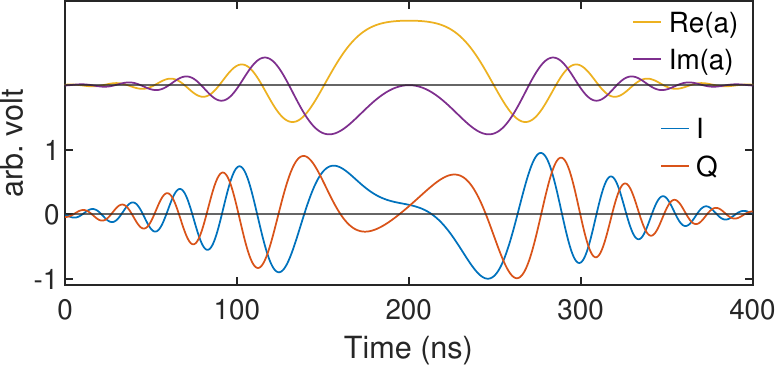}
    \caption{Chirped pulse for initial spin inversion, desired $\langle a (t) \rangle$ and corresponding $I/Q$ channels for optimal inversion.}
    \label{fig_methods:inversion_pulse}
\end{figure}

Starting from this state, the spins are inverted using an in-phase and quadrature modulated MW pulse. Similar to adiabatic fast passage methods for spins in free space, the starting point for the design of our inversion pulse is a chirped pulse of length $\qty[]{400}{\nano \second}$ with a Gaussian envelope, that covers a frequency interval of about -8 to +8 widths $\gamma_q$. As our spins are not in free space but strongly coupled to a cavity, we cannot use the chirped pulse directly but need to adapt it for this circumstance. Comparing the Maxwell-Bloch equations above with the optical Bloch equations, describing a two-level-system in free space driven by a classical coherent light field with driving amplitude $\Omega (t)$,
\begin{align}
         \dot{\sigma}_-  &= -\gamma_\perp  \sigma_- + \mathrm{i} \Omega(t)  \sigma_z \,,\label{eq_OBE_sigma_m} \\[5pt]
         \dot{\sigma}_z  &=  2\mathrm{i} \left(  \Omega(t)   \sigma_-  -  \Omega^\ast\!(t) \sigma_+   \right), \label{eq_OBE_sigma_z}
\end{align}
we can see that the role of the driving amplitude in the case of the coupled cavity-spin system is taken by the term $g_0  a^\dagger$. Disregarding $g_0$ as a proportionality constant that has to be determined experimentally, we can assume a desired photonic amplitude $a (t) = a_\mathrm{R} + \mathrm{i} a_\mathrm{I}$ given by the aforementioned chirped pulse and numerically solve for MW drive $\eta (t) = I - \mathrm{i} Q$ necessary to create it in the cavity (see Fig.~\ref{fig_methods:inversion_pulse}). Strictly speaking, this approach produces the correct effective inversion drive only for the spins with the center frequency, as only these are fully resonant. For other frequencies, the cavity acts as a filter and reduces the amplitude of the drive. Nevertheless, as the chirped pulse comes with a certain robustness to amplitude deviations, we still get a useful inversion efficiency for the whole spin ensemble. 

In the experiment, we scan the inversion pulse power to select the value for optimum inversion, which is in turn visible as the highest SR decay maximum for a given hold time. A similar scan is done to determine the optimum time for triggering the detuning via the current loop. The full inversion efficiency of the process is $(p_\mathrm{max} + 1)/2 \simeq 67 \%$ with a maximum value of $p$ taken from the simulation results in Fig.~3(a) in the main text.

\section{Delay time of the superradiant emission}
To derive \comment{Eq.~\eqref{eq_td}}Eq.~(3) in the main text, we neglect the inhomogeneous broadening $\Delta_s^j \approx 0$ and describe the spin ensemble as a giant dipole using $S_{\pm} = \sum_j \sigma_{\pm}^j$ and $S_{z} = \frac{1}{2} \sum_j \sigma_{z}^j$. When inserting Eq.~\eqref{eq_MBE_a} into Eq.~\eqref{eq_MBE_sigma_z} on resonance $\Delta_\mathrm{c} = 0$, we get
\begin{equation}
    \dot{S}_z = - \frac{2 g_0^2}{\kappa} S_+ S_-   -   \frac{\mathrm{i}g_0}{\kappa} (S_+ \dot{a} - \dot{a}^\dagger S_-) \,,
\end{equation}
where we now can neglect the second term as it is of lower order in the number of spins $N \propto S_-$. Effectively, the cavity acts similar to a vacuum environment for the SR burst, although enhancing the coupling of individual spins to the electromagnetic field. Its ability to store photons becomes of importance only at later times, when the cavity amplitude shows revivals, with excitations oscillating back and forth between cavity and spins. Then, we evaluate \cite{angerer2018superradiant}
\begin{equation}
    \begin{aligned}
        \langle \dot{S}_z \rangle &= - \frac{2 g_0^2}{\kappa} \langle S_+ S_- \rangle \\
        &= - \frac{2 g_0^2}{\kappa} \left(S + \langle S_z  \rangle\right)\left(S - \langle S_z \rangle +1\right)\,.
    \end{aligned}
\end{equation}
Now, we assume the giant dipole to be in a defined state $\lvert S,M \rangle$. By parametrizing $M=\langle S_z \rangle = \cos(\theta) N/2 $ with a tipping angle $\theta$ and using $S = N/2$, we can now solve for the delay time $t_\mathrm{D}$ where $\langle S_z \rangle = 0$, as the SR emission reaches its maximum when the giant dipole points to the equator. The resulting expression \cite{RAMII,superradiance}
\begin{equation}
    t_\mathrm{D} = t_0 - \frac{\kappa}{2 g_0^2 N} \log \left( \tan^2 \left( \frac{\theta}{2} \right) \right)
    \label{eq_td_methods}
\end{equation}
already resembles the one given in the main text. Now we linearize $\tan \theta \approx \theta$, neglect the constant offset $t_0$ and summarize the prefactor as $T_\mathrm{R} = \frac{\kappa}{2 g_0^2 N}$, representing the timescale of the SR emission process. We find good qualitative agreement of Eq.~\eqref{eq_td_methods} with our results. Quantitatively, when using the explicit values for $\kappa$ and $g_0^2 N = g^2_\mathrm{coll}$, the timescale of the SR emission is underestimated due to the approximations involved (in particular, neglecting the inhomogeneous broadening).

\section{Delay time rescaling and phase correction of the SR decay pulses}
The data collected in all experimental runs has some variance in the SR decay amplitudes. We assume this variance comes mostly from the solenoid switch located at the \qty[]{1}{\kelvin} stage which is used to disconnect the the pump line from port 1 of the cavity after the inversion pulse. The switch opens and closes a mechanical connection with a latching mechanism in the pump line used for the inversion pulse, thus leading to slightly different initial ensemble inversions between the runs. 

\begin{figure}[h]
    \centering
    \includegraphics[width=75mm]{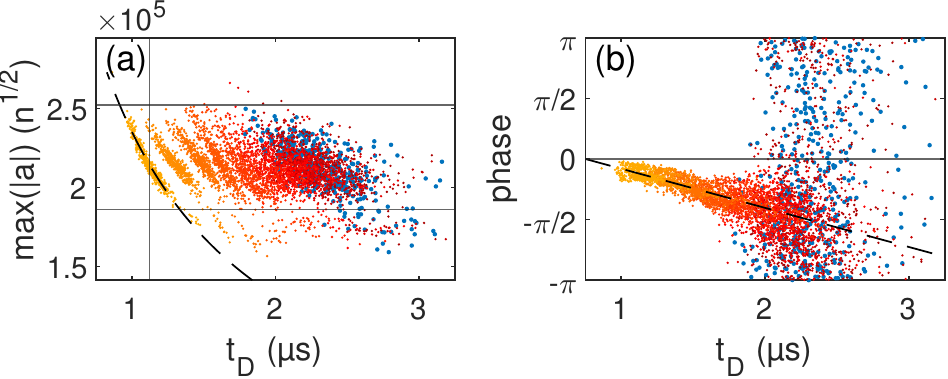}
    \caption{(a) All recorded data points of the SR decay amplitude maxima $\max(\lvert a \rvert) = \sqrt{ I_\mathrm{D}^2 + Q_\mathrm{D}^2}$ and phases $\varphi = \arctan(Q_\mathrm{D}/I_\mathrm{D})$ over the delay time $t_\mathrm{D}$, with the same color scheme as in the main text. The dashed line is a fit of the expected functional dependency to the highest power probe pulse values. (b) Phase drift over delay time with a linear fit to the six highest probe pulse powers. This drift is corrected by a linear shift that aligns the dashed line with the horizontal axis.}
    \label{fig:postsel}
\end{figure}

The initial inversion $ S_z $ is what determines the length of the $ S_- $ component during the SR decay process as the spin vector rotates towards the equator of the Bloch sphere. This in turn directly determines $\max(\lvert a \rvert)$ [see Eq.~\eqref{eq_MBE_a}]. The $N$ appearing in Eq.~\eqref{eq_td_methods} in this context parametrizes the initial inversion $ S_z $ [see derivation of Eq.~\eqref{eq_td_methods}\ above]. We can therefore use $ N \propto S_z \propto \max(\lvert a \rvert)$ and, inserting that into Eq.~\eqref{eq_td_methods}, we write more explicitly \mbox{$t_\mathrm{D} = t_0 + C/\max(\lvert a \rvert)$} for a fixed tipping angle. We confirm this dependency by fitting the data for the highest power probe pulses in Fig.~\ref{fig:postsel}(a). 

This systematic dependency of the delay time on the amplitudes would drastically broaden out the narrow distributions of the measured $t_\mathrm{D}$, if not accounted for. Therefore we adjust for this dependency by applying the same transformation to every measured $t_\mathrm{D}$, which aligns the dashed curve plotted in Fig.~\ref{fig:postsel}(a) with the vertical line at $t_\mathrm{D} \approx \qty{1.1}{\micro \second}$, the mean delay time value of the highest probe pulse power used in the experiment. The so transformed delay times are then used in Figs.~2(b) and (e) of the main text and referred to as rescaled $t_\mathrm{D}$. We also exclude the extreme outliers of $\max(\lvert a \rvert)$ from the analysis in the main text, which lie outwards of the two horizontal lines in Fig.~\ref{fig:postsel}(a).

The linear shift of the SR decay phase with $t_\mathrm{D}$ [see Fig.~\ref{fig:postsel}(b)] comes from a slight detuning between cavity and spins and is corrected for in the data presented in the main text for clarity. This phase correction is completely independent from the delay time rescaling.

\section{Distribution functions for delay time $t_\mathrm{D}$ and phase $\phi$ of the SR decay}
As described in the main text, the initial state of the inverted collective spin vector is located near the north pole of the Bloch sphere, close to the $z$-axis. Now, we approximate the surface near the north pole as a plane. Prior to the trigger pulse acting on the spin vector, the tipping angle $\theta = \arctan(\lvert S_- \rvert/S_z)$ is centered around  $\theta = 0$ but with a finite width of $\overline{\theta}$. After the trigger pulse displaces the spin state, the tipping angle follows the Rician distribution \cite{rice1945mathematical}
\begin{equation}
    f_\Theta(\theta,\eta,\overline{\theta}) = 
    \frac{\theta}{\overline{\theta}^2}\, \exp\!\left(-\frac{1}{2} \left(\frac{\theta^2}{\overline{\theta}^2} +\eta^2\right)\right) I_0\left(\frac{\theta \eta}{\overline{\theta}}\right),
    \label{eq_ftheta}
\end{equation}
with the modified Bessel function of the first kind $I_0$. The parameter $\eta$ expresses the displacement of the initial spin vector away from the origin in units of the width parameter $\overline{\theta}$, which we assume to be in the direction $\phi = 0$. This displacement is a result of the spin rotation caused by the trigger pulse. For $\eta \gg 1$ the distribution $f_\Theta$ becomes a Gaussian with mean value $\langle \theta \rangle = \eta \overline{\theta}$ and variance $\mathrm{Var}(\theta) = \overline{\theta}^2$. 
For the sake of completeness we also show the angular distribution for $\phi = \arg(S_-)$ of the resulting in-plane vector, which is given by \cite{angularmarginalization}
\begin{equation}
    \begin{aligned}
        f_\Phi(\phi,\eta) &= \frac{\eta}{\sqrt{2\pi}} \tilde{\varphi}(\eta) \left(1 + \eta\cos(\phi)\frac{\tilde{\Phi}(\eta\cos(\phi))}{\tilde{\varphi}(\eta\cos(\phi))} \right),
    \end{aligned}
    \label{eq_fphi}
\end{equation}
with the standard normal distribution $\tilde{\varphi}$ and its cumulative distribution function $\tilde{\Phi}$. As $\eta$ increases, the initially randomly distributed angle $\phi$ becomes more and more well defined and approaches $\phi = 0$.

We can infer the initial tipping angles $\theta$ from the delay times $t_\mathrm{D}$ using a simplified expression for the delay time derived above
\begin{equation}
    t_\mathrm{D} = - 2 T_\mathrm{R} \log\left(\frac{\theta}{2}\right),
    \label{eq_td}
\end{equation}
depending only on $\theta$ and $T_\mathrm{R}$. By applying a change of variables we arrive at the distribution for the delay times $t_\mathrm{D}$
\begin{equation}
    f_{t_\mathrm{D}}(t_\mathrm{D},\eta,\overline{\theta}) = f_\Theta\left(\theta(t_\mathrm{D},T_\mathrm{R}),\eta,\overline{\theta}\right) \left| \frac{d\,\theta(t_\mathrm{D},T_\mathrm{R})}{d\,t_\mathrm{D}} \right|.
    \label{eq_ftd}
\end{equation}

\section{Microwave setup}
For the generation of our MW inversion pulses, we use an arbitrary waveform generator to modulate the $I/Q$-quadratures onto a carrier wave created by a power source generator (PSG) at the cavity frequency of $\qty[]{3.1}{\giga \hertz}$. The pulses are gated using a fast MW switch, pass through a chain of digital attenuators and are amplified using a high power amplifier ($+\qty[]{40}{dB}$), before they enter the pump MW line, leading into the cryostat. At the $\qty[]{1}{K}$ stage inside the cryostat there is a relay switch, which can be used to completely decouple the pump line from the lower stages, blocking the room temperature thermal photons and the amplifier noise, which takes about $\qty[]{1}{\milli \second}$. \\
The probe pulses are created with another PSG and gated using a fast MW switch. Subsequently they pass through a second chain of variable digital attenuators, after which they are sent through the probe-line. In the experiment the probe-line has a fixed attenuation of $\qty[]{-69.5}{dB}$, of which $\qty[]{-20}{dB}$ are located right outside the cryostat, the rest distributed among the stages. The probe-line is connected to cavity port 2 using a splitter, together with the out-line.\\
Following the out-line upwards, we have two MW isolators with a combined isolation of $\qty[]{-20}{dB}$ and a \qty[]{-10}{dB} attenuator, for reducing thermal noise photons from the higher stages, before the signal is amplified with a low noise cryogenic amplifier. The signal is then demodulated using a homodyne detection setup, with the demodulation frequency supplied by the probe PSG. The two quadrature channels are finally measured with a high-speed data-acquisition system.

\section{Estimating the number of photons}
To estimate the number of photons contained in a probe pulse we do a calibration measurement of the attenuation $A_2 = \qty[]{-54.5}{\decibel} + \qty{5}{\decibel}$ at room-temperature for the probe-line inside the fridge up to port 2 of the cavity, where a value of $+\qty{5}{\decibel}$ is added to account for the decreased resistance of the lines when cold. Then we determine the MW power for the strongest probe pulses of the signal that enters the probe-line outside the fridge using a power spectrum analyzer, $P_{\mathrm{max}} = \qty[]{-58}{\micro\watt}$. The other probe pulses used in the experiment have variable attenuation decrements of \qty[]{-5}{\decibel} each, so the photon numbers change accordingly down to \qty[]{-45}{\decibel} relative to the highest power value.

\begin{figure}[]
    \centering
    \includegraphics[width=65mm]{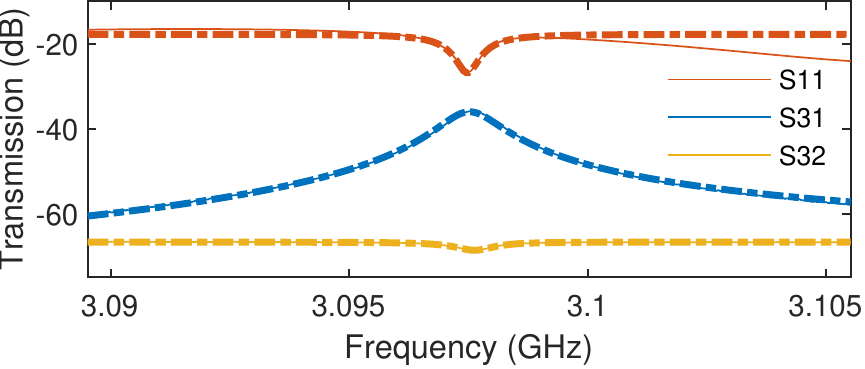}
    \caption{$S$-parameter traces (solid lines) of the cavity with far detuned spins where the numbers 1, 2 and 3 correspond to the pump, probe, and out-line and their respective fits (dashed lines). For these measurements an additional \qty[]{-20}{\decibel} attenuator at the probe-port entry was removed. This does not change the values for $\kappa_{1,2}$ obtained by fitting the dips, as they appear only relative to the base level. The parameter $\kappa_\mathrm{tot}$ manifests itself in the HWHM of the Lorentzian peak of $S_{31}$.}
    \label{fig:traces}
\end{figure}

Next, we determine the values of $\kappa_1$ and $\kappa_2$, the external coupling rates at both ports. For that, we measure the $S$-parameters of our system on resonance at $\qty[]{25}{\milli \kelvin}$ with the spins far detuned using the VNA as summarized in Fig.~\ref{fig:traces}.
By fitting the measured traces with the expected results from cavity input-output theory (reproduced for $\Delta_c = 0$, i.e., on resonance condition)  \cite{gardiner2004quantum} we obtain $\kappa_1$ and $\kappa_2$.
\begin{equation}
    \begin{aligned}
    \lvert S_{11} \rvert^2 &= A_1^2\left(2 \kappa_1/\kappa_\mathrm{tot} - 1 \right)^2 \\
    \lvert S_{31} \rvert^2 &= A_1 A_3 \left( 2 \sqrt{\kappa_1 \kappa_2}/ \kappa_\mathrm{tot}\right)^2 \\
    \lvert S_{32} \rvert^2 &= A_2 A_3 \left( 2 \kappa_2 / \kappa_\mathrm{tot} - 1 \right)^2 \\
    \end{aligned}
\end{equation}
Here, the subscripts in $A_{1,2,3}$ refer to the fixed MW line attenuations inside the cryostat for pump, probe, and out-line, respectively.

The time dependent cavity amplitude $|a|$ can be determined by solving the differential equation \eqref{eq_MBE_a} \mbox{(ignoring the spin term involving $g_0$)} and assuming a constant drive $\eta_d = \sqrt{ 2 \kappa_\mathrm{in} P_\mathrm{in}/ \hbar \omega_c }$ that starts at time $t=0$ as
\begin{equation*}
    a(t) = \frac{\eta_d}{\kappa_\mathrm{tot}}\left(1 - e^{-t \kappa_\mathrm{tot}} \right),
\end{equation*}
with the incident power $P_\mathrm{in}$ and the appropriate port's coupling rate $\kappa_\mathrm{in}$. 

In a pulse injected via port 2, with duration $\Delta t = \qty[]{100}{\nano \second}$, as used in the experiment, the number of photons in the cavity charges up to a maximum value of

\begin{equation*}
    n_{\mathrm{trig}}^{\mathrm{min}} = \frac{ P_\mathrm{max}^{\text{-}45\mathrm{dB}} } {\hbar \omega_c} A_2 \frac{2 \kappa_2}{\kappa_\mathrm{tot}^2}  \left( 1 - e^{-\Delta t \kappa_\mathrm{tot}} \right)^2 \approx 50,
\end{equation*}
where the power $P_\mathrm{max}^{\text{-}45\mathrm{dB}}$ measured outside is attenuated by an additional factor $A_2$ inside of the cryostat (c.f. Table~\ref{tbl:nphot}).

\begin{figure}[h]
    \centering
    \includegraphics[width=65mm]{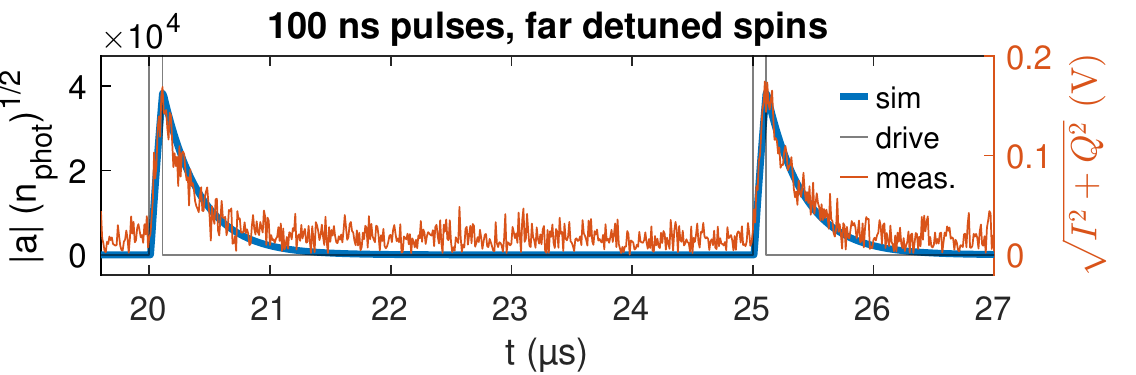}
    \vspace{-0.2cm}
    \caption{Pulse sequence and simulated cavity amplitude to calculate the $\mathrm{n}^{1/2}$ units from the voltages of the time-resolved $I/Q$ signal.}
    \label{fig:100ns_pulses_calib}
\end{figure}

For the pulse train measurements in the reduced cooperativity regime we use the same procedure to calculate the number of photons per $\qty[]{100}{\nano \second}$ pulse entering through the pump-line (c.f. Table~\ref{tbl:nphot_pulsetrain}),
\begin{equation*}
    n_{\mathrm{trig}}^{\mathrm{min}} = \frac{ P_\mathrm{MW} } {\hbar \omega_c} A_1 \frac{2 \kappa_1}{\kappa_\mathrm{tot}^2}  \left( 1 - e^{-\Delta t \kappa_\mathrm{tot}} \right)^2 \approx  \qty[]{1.5e9}{}.
\end{equation*}
These pulses are also used to calculate the $\lvert a \rvert$ units given as the square root of the cavity photon number $\mathrm{n}^{1/2}$ from the voltages of the time-resolved $I/Q$ measurement, see Fig.~\ref{fig:100ns_pulses_calib}.

Lastly, we calculate an estimate for the number of thermal photons in the cavity, when the solenoid switch at the \qty[]{1}{\kelvin} stage is open to decouple the higher temperature stages. We use the values of Table~\ref{tbl:attnstages} and evaluate according to
\begin{equation}
    \begin{aligned}
        n_i &= \overline{n}(T_i) + A_{i\,\text{-}1,i} \, n_{i\,\text{-}1} \, , \\    
        \overline{n}(T) &= \frac{1}{\exp(\hbar\omega_c/k_\mathrm{B} T )-1} \, ,
    \end{aligned}
\end{equation}
going down the stages for all MW lines. The dominant contribution are thermal photons from the \qty[]{1}{\kelvin} stage of the pump-line, which result in a value of $\overline{n}\approx 3$ photons.

We would like to emphasize that the calculated number of photons is dependent on the actual attenuation of the MW lines in the cryostat, which decreases from the room-temperature values when cooled. To account for this, we have estimated a $+\qty{5}{\decibel}$ change due to the temperature effects, which agrees well with the estimated number of decaying spins as shown in Fig.~1(d) in the main text. However, we acknowledge that these values may be subject to a factor of two uncertainty.

\begin{table}[h]
    \caption{Summary of the parameters used to estimate the number of photons entering the cavity via the $\qty[]{100}{\nano \second}$ trigger pulses through the probe-line.}
    \begin{ruledtabular}
    \begin{tabular}{cccc}
    $\kappa_2/2\pi$           & $\kappa_\mathrm{tot}/2\pi$  & $A_2$                     & $P^{\text{-}45\mathrm{dB}}_{\mathrm{max}}$        \\ \hline
    $\qty[]{59}{\kilo\hertz}$ & $\qty[]{586}{\kilo\hertz}$  & $\qty[]{-49.5}{\decibel}$ & $\qty[]{1.83}{\nano\watt}$  \\
    \end{tabular}
    \label{tbl:nphot}
    \end{ruledtabular}
\end{table}

\begin{table}[h]
    \caption{Parameters used to estimate the number of photons per pulse in the pulse sequences injected via the pump-line. This experiment was done in another cool-down of our cryostat, so the Q-factor of the resonator, therefore the $\kappa_{tot}$ value, exhibits some deviations from the ones above.}
    \begin{ruledtabular}
    \begin{tabular}{cccc}
    $\kappa_1/2\pi$         & $\kappa_{tot}/2\pi$ &  $A_1$                     & $P_\mathrm{MW}$          \\ \hline
    $\qty[]{182}{\kilo\hertz}$ & $\qty[]{516}{\kilo\hertz}$ & $\qty[]{-10.6}{\decibel}$ & $\qty[]{2.1}{\micro\watt}$   \\
    \end{tabular}
    \end{ruledtabular}
    \label{tbl:nphot_pulsetrain}
\end{table}

\begin{table}[h]
    \caption{Temperatures of the various stages inside the dilution refrigerator and corresponding attenuations (in the direction of lowering temperatures) between the respective stages to estimate the number of thermal cavity photons, when the solenoid switch at the nominal \qty[]{1}{\kelvin} stage is disconnected.}
    \label{tbl:attnstages}
    \vspace{-2.4mm}
    \rule{\columnwidth}{0.4pt} \\
    \vspace{-2.85mm} \rule{\columnwidth}{0.4pt}
    \begin{tabular}{c|c|c|c|c|c|c}
         \makebox[15.5mm]{\centering stage $i$} & 1 & 2 & 3 & 4 & 5 & 6 \\
        \hline
       $T_{i}$ (K)   & \makebox[10mm]{296} & \makebox[10mm]{42} & \makebox[10mm]{4} & \makebox[10mm]{0.9} & \makebox[10mm]{0.12} & \makebox[10mm]{0.025} \\ \hline
    \end{tabular}
    \begin{tabular}{c|ccccc}
        $A_{i,i+1}$ (dB) &  &  &  &  & \\ \hline 
        pump  & \multicolumn{1}{c|}{\makebox[10mm]{--}} & \multicolumn{1}{c|}{\makebox[10mm]{--}}  & \multicolumn{1}{c|}{\makebox[10mm]{--}} & \multicolumn{1}{c|}{\makebox[10mm]{-1.5}}  & {\makebox[10mm]{-2}} \\ \hline
        probe & \multicolumn{1}{c|}{-1.5} & \multicolumn{1}{c|}{-21.5} & \multicolumn{1}{c|}{-1.5} & \multicolumn{1}{c|}{-11.5} & -13.5 \\ \hline
        out   & \multicolumn{1}{c|}{-1.5} & \multicolumn{1}{c|}{-1.5}  & \multicolumn{1}{c|}{-1.5} & \multicolumn{1}{c|}{-1.5}  & -30
    \end{tabular}
    \rule{\columnwidth}{0.4pt} \\
    \vspace{-2.85mm} \rule{\columnwidth}{0.4pt}
\end{table}

\section{$T_1$ measurements using the Vector Network Analyzer}
\begin{figure}[b]
    \centering
    \includegraphics[width=85mm]{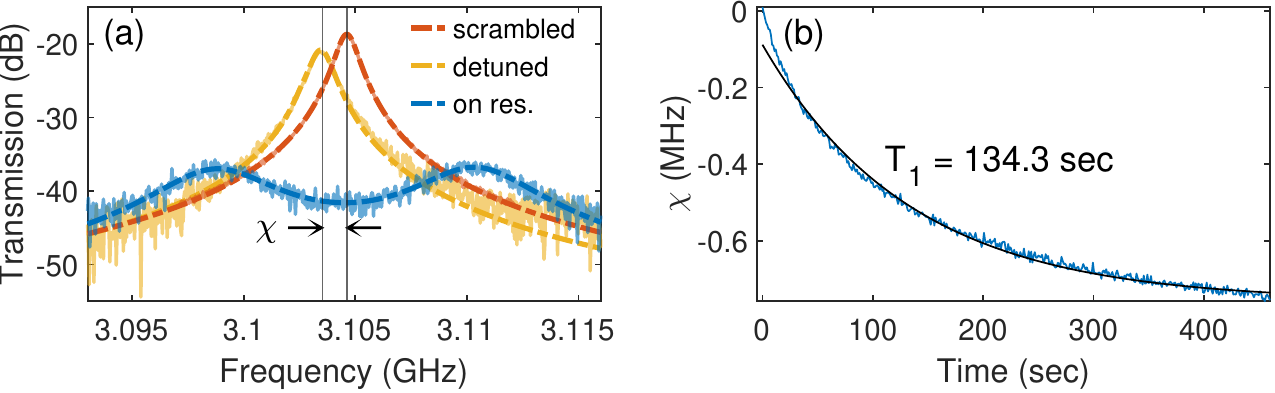}
    \caption{(a) VNA transmission measurement of the hybrid system in its ground state on resonance (blue), with the detuning loop off (spins detuned, yellow) and measured using high input power to scramble the spins (red), together with respective fits of the steady state transmission. (b) Dispersive shift $\chi$ over time, extracted from Lorentzian fits to the transmission data. The black fit line corresponds to a simple exponential decay law.}
    \label{fig:vnaT1}
    \vspace{3mm}
\end{figure}
In the main text we discuss a fast relaxation of the stored spin inversion with a characteristic timescale of $\qty[]{7.6}{\milli \second}$. This observed fast decay is contrasted by a slow relaxation from a randomized spin ensemble, as shown in Fig.~\ref{fig:vnaT1}. The initial state for measuring this slow relaxation is created by repeatedly sweeping across the resonance with the VNA using a high input power for \qty[]{30}{\second}. This way, we scramble the spins, creating a state with spin polarization zero. For a large ensemble detuning $\delta \gg g_\mathrm{coll}$ the dispersive shift $\chi$, represented graphically in Fig.~\ref{fig:vnaT1}(a), allows a direct way to determine the long $T_1$ time of the spins with the result

\begin{equation}
    \chi(t) = \frac{g_\mathrm{coll}^2}{\delta} \langle S_z (t) \rangle \,,
\end{equation}
as similarly employed in \cite{astner2018solid}.

\section{Nitrogen vacancy center spins and diamond sample}
The spin ensemble used in this work consists of negatively charged nitrogen vacancy centers in diamond (NV), which are made up of a substitutional nitrogen atom with an adjacent lattice vacancy. This \mbox{paramagnetic} impurity has an electron spin $S= 1$ and can be described by the Hamiltonian $H = \hbar D S_z^2 + \mu B S$, with the zero field splitting $D / 2\pi = \qty{2.88}{\giga \hertz}$ and $\mu / 2\pi = \qty[per-mode=symbol]{28}{MHz \per mT}$. The diamond symmetry results in four possible orientations of the NV centers. \\

The roughly cube shaped diamond samples with side length $d \sim \qty[]{200}{\micro \meter}$ were cut from a larger sample by Delaware Diamond Knives. The larger sample was created similarly to the one characterized in detail in \cite{astner2018solid}, referred to as ``N1" therein. It was made by irradiating a commercially available high-pressure high-temperature diamond with an initial nitrogen concentration of 200 ppm and naturally abundant ${}^{13}\mathrm{C}$
isotopes with our in-house neutron source (TRIGA Mark II reactor) for lattice vacancy creation. It was irradiated with a fluence of $5 \times 10^{17} \mathrm{cm}^{-2}$ for 50 h and annealed at 900 \textdegree C  for 3 h.

The dimensions of the diamond sample are determined from microscope images. It has the approximate shape of a truncated pyramid with rectangular bottom and top faces of roughly $210 \times \qty{190}{\square \micro \meter}$, and $120 \times \qty{100}{\square \micro \meter}$ respectively, and a height of $\qty{210}{\micro \meter}$, resulting in a volume of $\qty{5.16e6}{\cubic \micro \meter}$. With the number of spins estimated from the measured collective coupling versus the simulated single spin coupling of the resonator (see section \ref{sec:resonator_sec}), we get a value for the density of NV centers of roughly 7 ppm.

\section{Superconducting double chip resonator}
\label{sec:resonator_sec}
\begin{figure}[t]
    \centering
    \includegraphics[width=85mm]{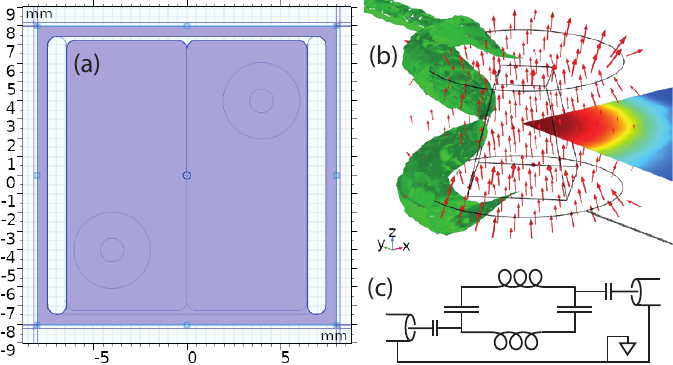}
    \caption{(a) Layout of the Niobium patterned chip, making up one half of the double chip resonator. The faint pairs of concentric rings along the diagonal show the placement of the capacitive coupling antennas, on opposite sides of the double chip assembly. (b) Three dimensional COMSOL simulation of the oscillating magnetic field (red arrows, and color plot inset showing $\lvert B_\mathrm{osc} \rvert$ in the XY-plane) between the center holes of the two chips. The sample, a truncated rectangular pyramid, lies inside the green isosurface (only one quadrant shown) enclosing the region in which the field strength deviates less than 5\% from the average value of $\lvert B_\mathrm{osc} \rvert$ in the sample. (c) Equivalent circuit diagram of the resonator. }
    \label{fig:resonator_sim}
\end{figure}

Our superconducting resonator design consists of two Niobium patterned chips, each acting as a split ring resonator \cite{splitringres}, with a layout detailed in Fig.~\ref{fig:resonator_sim}(a). The oscillating charges are stored on large capacitor pads, separated by the \qty{4}{\micro \meter} slit that extends outwards from the center hole with a radius of $r = \qty{190}{\micro \meter}$. By stacking the chips on top of each other with a distance of $d = \qty{250}{\micro \meter}$ approximating the geometry of plate capacitors, the overall capacitance of the structure is amplified. In order to magnetically couple to spins, a high capacitance in a lumped element resonator is desirable, which creates high currents (i.e. high magnetic fields) when the charge imbalance stored in the capacitors oscillates. The currents travel mainly along the perimeters of the center holes when oscillating from left to right during one half oscillation period, creating a homogeneous magnetic field in the central sample volume with a homogeneity of over 95\% [see Fig.~\ref{fig:resonator_sim}(b)]. The outer current path around the split ring structure on each chip [shown in Fig.~\ref{fig:resonator_sim}(a), but omitted in the main text for simplicity] is considerably longer and does not contribute much to the total inductance but is thought to help contain the fields closer to the resonator, thereby improving the $Q$ factor, as eddy currents in the copper box are suppressed. The resonator is capacitively coupled to the two MW ports with two antenna pins placed diagonally opposite above and below the chip stack, its equivalent circuit diagram being shown in Fig.~\ref{fig:resonator_sim}(c). 

In Table~\ref{tbl:resonator} we show simulation results for frequency $\omega_\mathrm{c}$, center magnetic field $B_0$, and mode volume $V_\mathrm{eff}$, using the finite-element simulation software \mbox{COMSOL}. We calculate the single spin coupling strength \mbox{$g_0 /2\pi = \gamma_\mathrm{NV} B_0^\mathrm{sim} \sqrt{2/3} \approx \qty{2.05}{\hertz}$}, using $\gamma_\mathrm{NV} /2\pi \approx \qty{28}{\mega \hertz \per \milli \tesla}$ and a geometric factor of $\sqrt{2/3}$ that accounts for the orientations of the NV center spins in the diamond with respect to the magnetic field direction.

Comparing the collective coupling strength as determined from the experiment with the single spin coupling allows us to estimate the number of spins as $N = g_\mathrm{coll}^2/g_0^2 \approx \qty{6.4e12}{}$.

The quality factor and resonance frequency of the resonator deviate slightly between the two cooldowns during this experiment with the values of $Q^\mathrm{I} = 2587,\quad Q^\mathrm{II} = 3010$, and $\omega_\mathrm{c}^\mathrm{I}/2\pi = \qty{3.098}{\giga \hertz},\quad \omega_\mathrm{c}^\mathrm{II}/2\pi = \qty{3.105}{\giga \hertz}$, respectively.

\begin{table}[h]
    \caption{Frequency, oscillating magnetic field at the center, and effective mode volume simulated using COMSOL.}
    \begin{ruledtabular}
    \begin{tabular}{ccc}
     $\omega_\mathrm{c}^\mathrm{sim}/2\pi$ & $B_0^\mathrm{sim}$ & $V_\mathrm{eff}^\mathrm{sim} = \int B^2 dV / {B_0^\mathrm{sim}}^2 $ \\ \hline
     $\qty{3.0}{\giga \hertz}$ & $\qty{8.97e-11}{\tesla}$ & $\qty{3.1e8}{\cubic \micro \meter}$ 
    \end{tabular}
    \end{ruledtabular}
    \label{tbl:resonator}
\end{table}

\comment{
We now attempt to calculate the strength of the resonator's oscillating magnetic field from geometric principles. By using the formula for a plate capacitor, it is possible to calculate the value of \mbox{$C_1 = \epsilon_0 (13 \times \qty{6.5}{\square \milli \meter})/\qty{250}{\micro \meter} \approx \qty{3}{\pico \farad}$} (see Fig.~\ref{fig:resonator}(a,b)). As the two equal capacitances $C_1$ are in series, the full capacitance of the circuit is $C = C_1/2$. 
Now, combining the frequency $\omega_\mathrm{c} = 1/\sqrt{L C} \approx \qty{3}{\giga \hertz} \times 2 \pi $ with the magnetic energy of the vacuum fluctuations in the inductor $\hbar \omega_\mathrm{c}/4 = L I^2 /2$ (its other half being stored in the capacitor), we can derive an expression for the oscillating current associated with the vacuum fluctuations $I_0 = \sqrt{\hbar \omega_\mathrm{c}^3 C /2} \approx \qty{23}{nA}$. Finally, we estimate the magnetic field in the center of the resonator using two circular current loops of radius $r$, separated by a distance $d$ carrying a current of $I_0$ (a common textbook example), arriving at $B_0^\mathrm{geom.} \approx \qty{8.9e-11}{\tesla} $.}

\section{Extended pulse train plots}
\begin{figure}[h]
    \centering
    \vspace{4mm}
    \includegraphics[width=80mm]{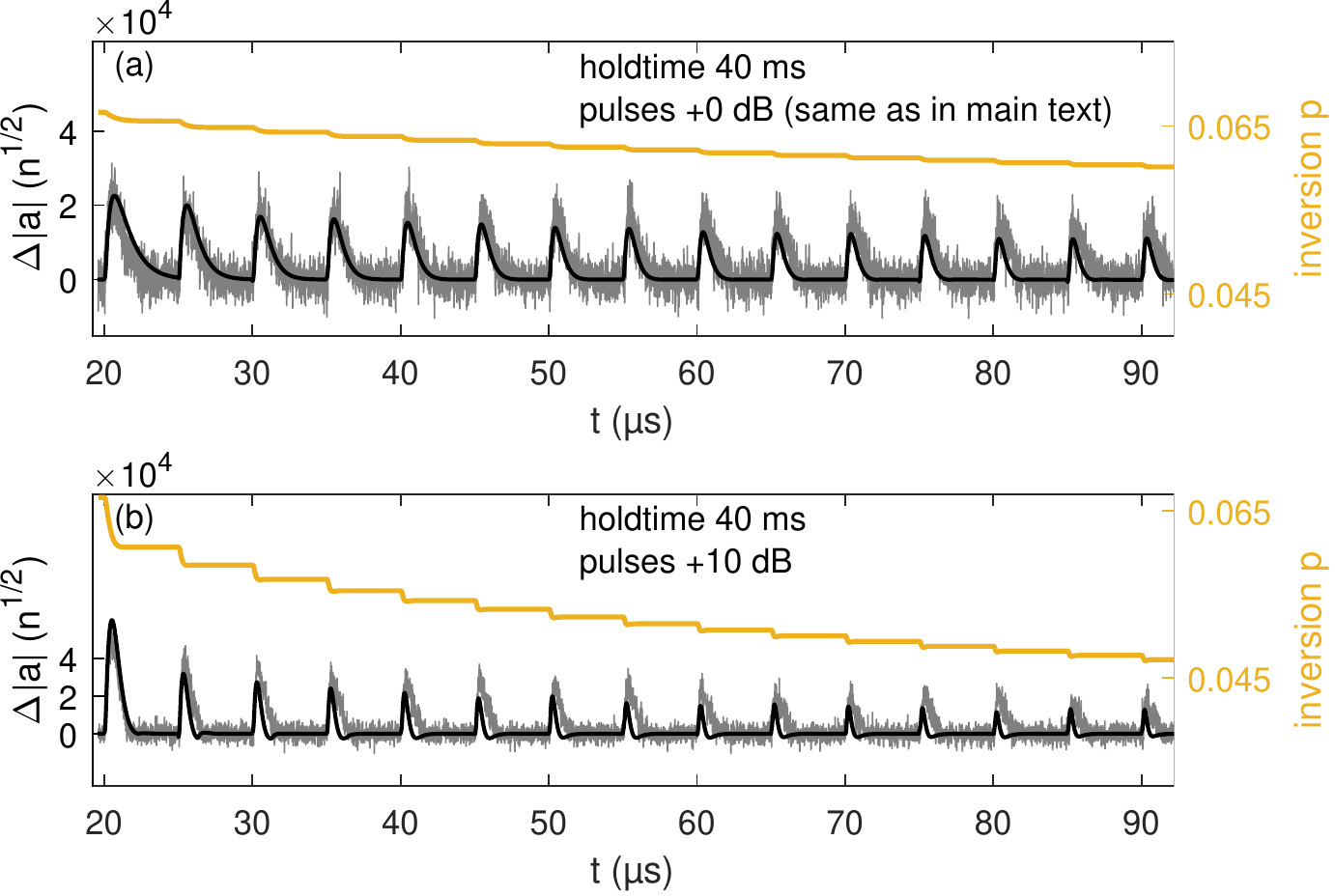}
    \caption{Amplification of $\qty{100}{\nano \second}$ pulses in the reduced effective cooperativity regime ($pC < 1$) for extended times using different pulse powers. To better visualize the amplification, we plot the difference $\Delta \lvert a \rvert$ of the measured amplitudes with partially inverted spins to the empty cavity signal (spins far detuned). The plot also includes a semi-classical model simulation of the inversion dynamics over time. Compare with Fig.~3(a) in the main text.}
    \label{fig:tens_injected}
\end{figure}

We show experimental data for the successive amplification of multiple MW pulses in the case of a partially inverted spin ensemble with reduced effective cooperativity $pC < 1$, as an extension of the measurements presented in the main text [see Fig.~\ref{fig:tens_injected} and compare with Fig.~3(a)]. Furthermore, we plot another measurement run using a pulse power that is $+\qty{10}{\decibel}$ higher than what is shown in Fig.~3(a) in the main text, both with the same hold time of $\qty{40}{\milli \second}$. For the stronger pulses, each pulse relaxes more of the stored inversion. We note that our semi-classical model can generally reproduce the measured data with a fair qualitative agreement but is not able to fit the pulse train measurements as well as in the case of the free decays [compare with Fig.~1(c,d) in the main text].

\bibliography{Triggered_Superradiance}

\end{document}